\documentclass[fleqn,usenatbib]{mnras}

\usepackage{newtxtext,newtxmath}

\usepackage[T1]{fontenc}

\DeclareRobustCommand{\VAN}[3]{#2}
\let\VANthebibliography\thebibliography
\def\thebibliography{\DeclareRobustCommand{\VAN}[3]{##3}\VANthebibliography}

\usepackage{graphicx}	
\usepackage{amsmath}	
\usepackage[normalem]{ulem}

\title[PAH 3.4 micron feature]{The PAH 3.4 micron feature as a tracer of shielding in the Orion Bar and NGC 6240}

\author[N. Thatte et al.]{
N. Thatte $^{1}$\thanks{E-mail: niranjan.thatte@physics.ox.ac.uk},
D. Rigopoulou$^{1,2},$
F. R. Donnan$^{1}$,
I. Garcia-Bernete$^{3}$,
M. Pereira-Santaella$^{4}$,
B. Draine$^{5}$
\newauthor
O. Veenema$^{1}$,
B. Kerkeni$^{6,7,8}$
A. Alonso-Herrero$^{3}$,
L. Hermosa Mu{\~n}oz$^{3}$,
G. Speranza$^{4}$
\\
$^{1}$ Department of Physics, University of Oxford, Keble Road, Oxford OX1 3RH, UK \\
  $^{2}$ School of Sciences, European University Cyprus, Diogenes street, Engomi, 1516 Nicosia, Cyprus\\
   $^4$ Instituto de F\'isica Fundamental, CSIC, Calle Serrano 123, 28006 Madrid, Spain\\
   $^{3}$ Centro de Astrobiolog\'{\i}a (CAB), CSIC-INTA, Camino Bajo del  Castillo s/n, E-28692 Villanueva de la Ca\~nada, Madrid, Spain\\
   $^5$ Department of Astrophysical Sciences, Princeton University, Princeton, NJ 08544-1001, USA\\
   $^{6}$De Vinci Higher Education, De Vinci Research Center,
92916 Paris, France\\
   $^{7}$ Facult\'e des Sciences de Tunis, Laboratoire de Physique de la Mati\`ere Condens\'ee, Universit\`e Tunis el Manar, Tunisia 2092\\
   $^{8}$ISAMM, Universit\'e de la Manouba, La Manouba 2010, Tunisia
 \\
}

\date{Accepted XXX. Received YYY; in original form ZZZ}

\pubyear{\the\year{}}

\begin{document}
\label{firstpage}
\pagerange{\pageref{firstpage}--\pageref{lastpage}}
\maketitle

\begin{abstract}
We have carried out a detailed analysis of the 3.4\,$\mu$m spectral feature arising from Polycyclic Aromatic Hydrocarbons (PAH), using JWST archival data. For the first time in an external galaxy (NGC 6240), we have identified two distinct spectral components of the PAH 3.4\,$\mu$m feature: a shorter wavelength component at 3.395\,$\mu$m, which we attribute to short aliphatic chains tightly attached to the aromatic rings of the PAH molecules; and a longer wavelength feature at 3.405\,$\mu$m that arises from longer, more fragile, aliphatic chains that are weakly attached to the parent PAH molecule. These longer chains are more easily destroyed by far-ultraviolet photons (>5eV) and PAH thermal emission only occurs where PAH molecules are shielded from more energetic photons by dense molecular gas.  We see a very strong correlation in the morphology of the PAH 3.395\,$\mu$m feature with the PAH 3.3\,$\mu$m emission, the latter arising from robust aromatic PAH molecules. We also see an equally strong correlation between the PAH 3.405\,$\mu$m morphology and the warm molecular gas, as traced by H2 vibrational lines.  We show that the flux ratio PAH~3.395/PAH~3.405 < 0.3 corresponds strongly to regions where the PAH molecules are shielded by dense molecular gas, so that only modestly energetic UV photons penetrate to excite the PAHs.  Our work shows that PAH 3.405\,$\mu$m and PAH 3.395\,$\mu$m emission features can provide robust diagnostics of the physical conditions of the interstellar medium in external galaxies, and can be used to quantify the energies of the photon field penetrating molecular clouds.

\end{abstract}




\section{Introduction}
Polycyclic aromatic hydrocarbon (PAH) molecules are ubiquitously present in the Interstellar Medium (ISM) of our Galaxy and of star-forming galaxies (e.g. \citealp{Smith2007, Galliano2008, mps10, Lai23, hernandez24, donnan23, rigo24}) and Active Galactic Nuclei (AGN) (\citealp{IGB22ers,IGB24pah, zhang24, ramosalmeida25}) locally and at high redshifts (e.g. \citealp{chen24, spilker23}). They exhibit distinct infrared (IR) features at 3.3, 6.2, 7.7, 11.3 and 17\,$\mu$m (e.g. \citealp{LegerPuget94}) which are attributed to thermal emission by PAH molecules upon absorption of UV photons from their surrounding radiation fields.

PAH molecules consist of carbon and hydrogen atoms arranged in benzene ring structures (e.g. \citealp{tielens08}). The emission bands are believed to originate from the stretching and bending of the vibrational modes of the C-H and C-C bonds (e.g. \citealp{allamandola85}) with the smaller PAH molecules reaching high excitation temperatures and emitting strongly in the C–H stretching region encompassing the 3-4\,$\mu$m regime (\citealp{ricca12}). PAHs are present in Photodissociation Regions (PDR) where they play a key role in heating the gas through the photoelectric phenomenon (e.g. \citealp{weindraine01}). Understanding the
nature, origin and evolution of PAHs across cosmic time and galaxy properties offers crucial insights into the interplay between galaxy evolution, dust, and gas.

The 3.3\,$\mu$m PAH emission band is known to be an excellent tracer of small,
neutral PAHs 
(e.g. \citealp{rigo21, rigo24, DraineLee07, Draine21}). 
In addition to the 3.3\,$\mu$m band, a weaker band can also be observed around $\sim$3.4\,$\mu$m. 
The exact relationship between the 3.4\,$\mu$m and 3.3\,$\mu$m bands is
unclear. The 3.4\,$\mu$m band is often assigned to aliphatic side
chains attached to PAH molecules (\citealp{duleywilliams81,
pauzat99}) although this idea has previously been disputed
(e.g., \citealp{Sandford91}).

The 3.4\,$\mu$m PAH band has been seen to exhibit a lot of
variability, depending on whether the carrier is purely aliphatic or attached to an aromatic ring system (\citealp{Wexler67}), and even
shows some variability depending on the specific ring system
to which it is attached, and the configuration of the resulting alkyl PAH
molecule (e.g. \citealp{Yang13}). Significant variation is also seen
in astrophysical sources, where some planetary nebulae show a 3.4\,$\mu$m band which is stronger than the 3.3\,$\mu$m band (\citealp{Hrivnak07}), despite it normally being the weaker of the two emission
features. Other observations suggest the possibility of the 3.4\,$\mu$m band arising from an anharmonicity effect (\citealp{Barker87,  LiDraine12}) or superhydrogenation (\citealp{sandford13, yang20}). In addition, \cite{Roche96} suggested that the feature is more prominent in nitrogen rich objects.
The comparative fragility of the aliphatic hydrocarbons is possibly the reason for the estimate of <15\% of the carbon
atoms responsible for emitting the PAH bands being in aliphatic
form (\citealp{LiDraine12}), however, this does not necessarily
apply to the 3.4\,$\mu$m carriers observed in absorption (\citealp{Chiar13}; \citealp{Pendleton25}).
In this paper, we will adopt the hypothesis that the 3.4\,$\mu$m band is  attributed to aliphatic hydrocarbons. 
We will focus, in particular, on understanding the nature of the carriers, the relation between the 3.3 and 3.4\,$\mu$m bands and the use of the latter 
in probing the local ISM conditions. 
So far, detection of the PAH 3.4\,$\mu$m band has been limited to galactic sources (e.g., \citealp{Roche96, JdM90, Ohsawa16}) and a handful of IR luminous galaxies (e.g., \citealp{imanishi10, Lai23}).

With the advent of JWST it is now possible to study PAH emission in a spatially resolved manner in a wide range of sources and environments. Observations with JWST$/$NIRSpec (covering the 1-5\,$\mu$m wavelength range) have paved the way for the study of the PAH 3.3 and 3.4\,$\mu$m emission features in galactic and extragalactic sources. 
Such observations 
have shown that the PAH 3.3\,$\mu$m and 3.4\,$\mu$m bands are observed in galaxies of the nearby
Universe (e.g., \citealp{Lai23, IGB24soda, Perna24}) and up to a redshift of four (the lensed
galaxy SPT0418-47; \citealp{spilker23}), where only the
3.3\,$\mu$m band is detected with JWST$/$MIRI MRS. 
Finally, the PAH 3.3 and 3.4\,$\mu$m emission features have also
been detected by JWST, through its Near Infrared Camera
(NIRCam), in a number of galaxies at intermediate redshifts
z $\sim$ 0.2–0.5 in the Great Observatories Origins Deep Survey-
South (GOODS-S; \citealt{Lyu25}).

In this work, we focus on the emission from the weaker 3.4 $\mu$m band and how the feature might be linked to the local physical conditions in the ISM of galaxies. We first investigate the nature of the aliphatic PAH 3.4\,$\mu$m emission band and then explore the conditions that might influence its presence and strength. In particular, we want to explore whether clouds of molecular hydrogen (H$_{2}$) might be shielding PAHs in strong UV radiation fields, as suggested by \citealp{rigo02, roussel07, AAH20}. For this purpose we employ archival observations of the Orion Bar, 
which offers unparalleled spatial resolution, and use this as a template to guide our investigations of nearby galaxies. In addition, we employ observations of the nearby Luminous Infrared Galaxy (LIRG) NGC~6240, A.~Alonso-Herrero) 
which is well-known for its strong PAH and H$_{2}$ emission (e.g. \citealp{rigo99, Lutz03, AAH14, Hermosa25}). 
The article is structured as follows: in Sect. 2, we give a brief description of the targets and the ensuing data analysis. In section 2.1.2 we discuss in detail the procedure we used for fitting and extracting the PAH3.3 and PAH 3.4 features.
In Sect. 3 we present theoretical spectra of molecules containing aliphatic side-groups, computed using Density Functional Theory (DFT). In Sect. 4 we analyse the spatial distribution of the features with a detailed discussion in Sect 5. Finally we present our conclusions in Sect. 6.
Throughout the paper we use H$_{0}$ = 70 km s$^{-1}$ Mpc$^{-1}$, 
$\Omega_{m}$ = 0.3, and $\Omega_{\Lambda}$ = 0.7, unless otherwise stated.

\section{Observations and Data Reduction}
\label{sec:obsData}
\subsection{JWST Data}
We used archival JWST NIRSpec IFU Spectroscopy to study the PAH emission features in the 3.2--3.6\,$\mathrm{\mu m}$ wavelength range. Data for NGC~6240 were obtained as part of the Guaranteed Time Observations (Program ID 1265, P.I.~A.Alonso-Herrero) in the F290LP-G395H configuration.  Data for the Orion Bar were obtained as part of the Early Release Science (ERS) programme (Program ID 1288, P.I.~O. Berné), also with the F290LP-G395H configuration of NIRSpec in integral field mode. 
The NGC~6240 
data were calibrated and reduced with the standard JWST NIRSpec pipeline, with one exception. We added an extra step between Stages 2 and 3 of the pipeline, correcting for residual bad, hot and dead pixels that had not been flagged by the pipeline, with a procedure similar to that described by \cite{IGB24soda}. 
For the Orion Bar, we used the Stage 3 data products from the MAST archive without any alterations.

\subsubsection{NGC~6240 analysis}
NGC\,6240 is located at a distance of 108\,Mpc, so each 0\farcs1 IFS spaxel corresponds to 52\,pc (NASA/IPAC Extragalactic Database). The aims of the data analysis were to recover the integrated flux 
in the PAH features in the range 3 to 4\,$\mu$m 
as well as a number of emission lines of molecular and ionized Hydrogen.  The hydrogen recombination lines have been analysed and presented by \cite{Ceci25} whereas an analysis of the molecular H$_{2}$ lines has been presented in  \cite{Hermosa25} for NIRSpec and JWST$/$MIRI datasets, respectively. For the purposes of this work, we used the H$_2$\,0$-$0\,S(8) line of molecular Hydrogen (rest wavelength 5.053\,$\mathrm{\mu m}$), and the Pfund\,$\gamma$ transition of ionized Hydrogen (rest wavelength 3.740\,$\mathrm{\mu m}$). We chose Pf\,$\gamma$ over Pf\,$\beta$, as the Pf\,$\beta$ transition is harder to isolate from nearby CO features, particularly in the nuclear regions. We also used H$_2$\,0$-$0\,S(8) instead of H$_2$\,0$-$0\,S(9) for the same reason. No correction for differential extinction was applied, given 
the relative proximity of the wavelengths of the features examined. 

As described by \cite{Ceci25} and \cite{Hermosa25}, NGC~6240 exhibits very strong molecular hydrogen emission throughout its nuclear region, and relatively modest ionized hydrogen emission.  \cite{Ceci25} present a detailed kinematic analysis of several emission lines in the nuclear region of NGC~6240, including the H$_2$\,1$-$0\,S(1) line. They show that the molecular hydrogen emission exhibits two components in many spaxels within the NIRSpec IFU field-of-view, as do the ionized hydrogen features.  Consequently, we used a double Gaussian fit (and a linear continuum) for all molecular hydrogen and ionized hydrogen lines in NGC~6240, using the {\tt mpfit2gauss} routine. The {\tt mpfit} library provides Levenberg-Marquardt least squares minimisation fits to user specified functions, we used it throughout our analysis for Gaussian and Drude profile fitting, with appropriate constraints.

\subsubsection{PAH features}{\label{obs_pah}}

\begin{figure}
	\includegraphics[width=0.95\columnwidth,angle=180]{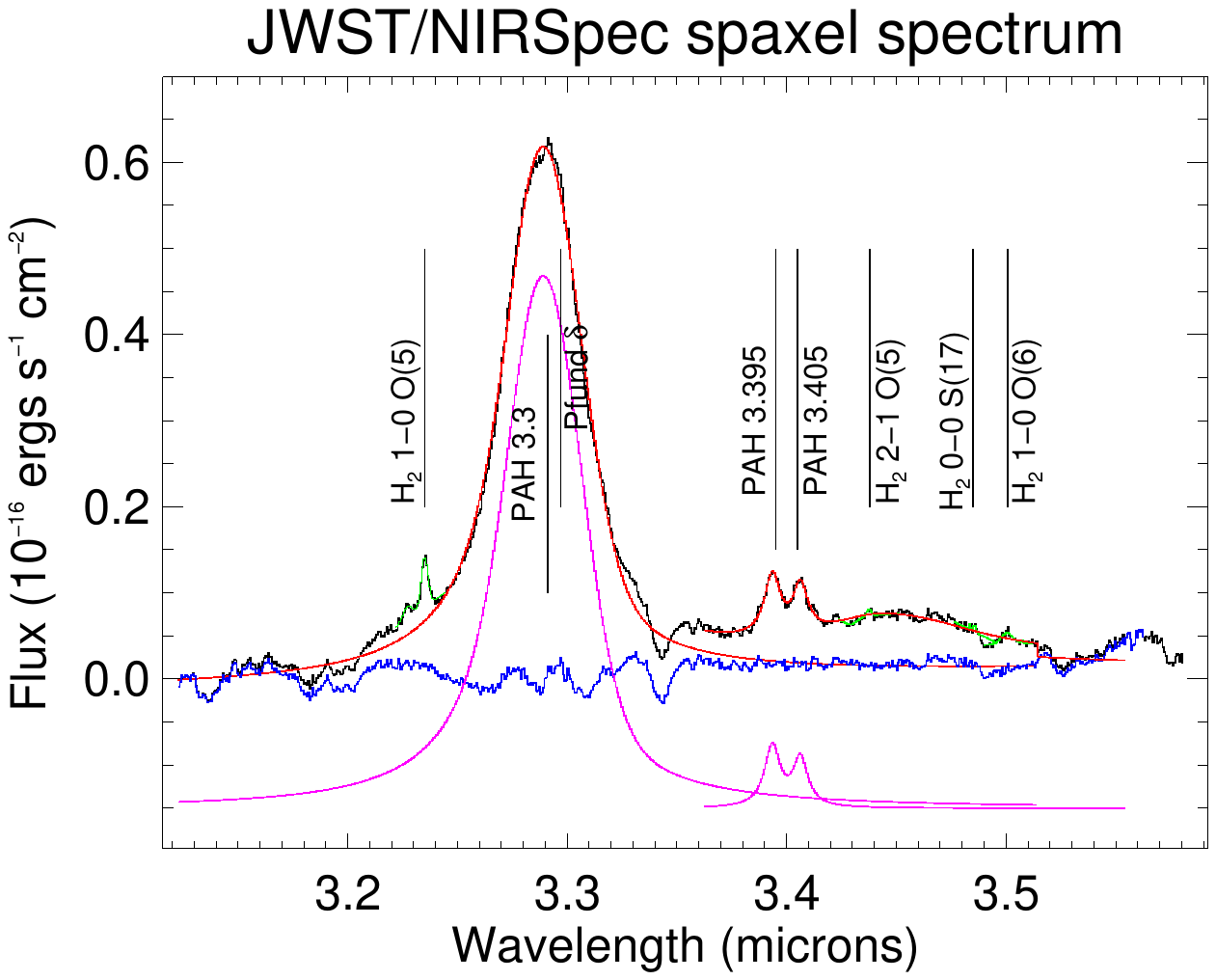}
    \caption{Rest-frame spectrum of a representative single spaxel in the nuclear region of NGC~6240 (0\farcs 1 E, 0\farcs 4 S of the southern nucleus). Plotted flux is per spectral pixel. Median 1--$\sigma$ error across the spectral range is 4.4\% of plotted flux. Observed continuum subtracted spectrum in black, fits to the PAH features in red, fits to H$_2$ features in green, post-fit residual in blue. PAH Drude profiles overlaid in magenta (shifted for clarity, see Sec. \ref{obs_pah} for details). }
    \label{fig:spaxelspec}
\end{figure}

Before fitting the PAH features, we fitted and subtracted the emission from a number of lines in the same wavelength range as the PAH emission, namely the H$_2$\,1$-$0\,O(5) line ($\lambda_\mathrm{rest} = 3.235\,\mathrm{\mu m}$), the H$_2$\,2$-$1\,O(5) line ($\lambda_\mathrm{rest} = 3.438\,\mathrm{\mu m}$), the H$_2$\,0$-$0\,S(17) line ($\lambda_\mathrm{rest} = 3.485\,\mathrm{\mu m}$), and the H$_2$\,1$-$0\,O(6) line ($\lambda_\mathrm{rest} = 3.500\,\mathrm{\mu m}$).  A spectrum of a typical spaxel in the nuclear region with fits to these lines overlaid in green, is shown in Figure \ref{fig:spaxelspec}.

We fit the 3.3\,$\mathrm{\mu m}$ PAH feature, using the combination of a Drude profile (\citealp{Smith2007}) and a Gaussian, plus a linear continuum. The Gaussian accounts for any Pf\,$\delta$ emission ($\lambda_\mathrm{rest} = 3.297\,\mathrm{\mu m}$) just slightly redward of the PAH\,3.3 peak, although the ionized hydrogen emission is weak compared to the PAH and H$_2$ features in the nuclear region of NGC~6240.

Figure \ref{fig:spaxelspec} shows that the 3.4\,$\mathrm{\mu m}$ PAH feature in NGC~6240 is clearly double-peaked. Although the double-peaked profile of the 3.4 $\mu$m feature has been previously observed in external galaxy spectra (e.g., \citealp{Lai23, spilker23}), the two components have not been distinguished as separate features, and have typically been fit using a single Drude profile.
For our analysis 
we follow the feature decomposition adopted by \cite{Peeters24} for the PAH features in the 3.37--3.6\,$\mathrm{\mu m}$ range in the Orion Bar and fit Drude profiles to five features centered at rest wavelengths of 3.395, 3.405, 3.425, 3.464 and 3.516\,$\mathrm{\mu m}$, with guess widths as per Table\,H.1 of \cite{Peeters24}.
The fit includes a linear continuum spanning the wavelength range, and constraints on the 3.395 and 3.405\,$\mathrm{\mu m}$ features to be blueward and redward of 3.4\,$\mathrm{\mu m}$ respectively.

\subsubsection{Orion Bar analysis}
\begin{figure*}
\includegraphics[width=0.5\textwidth]{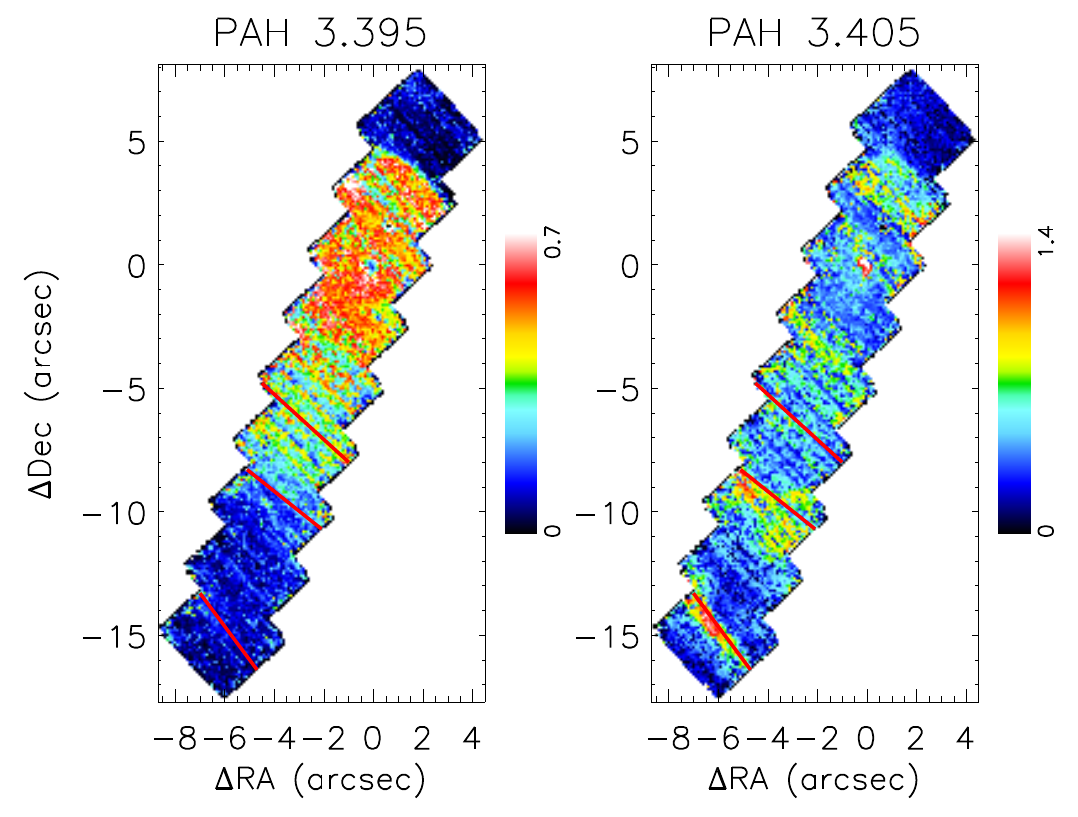}
\includegraphics[width=0.47\textwidth]{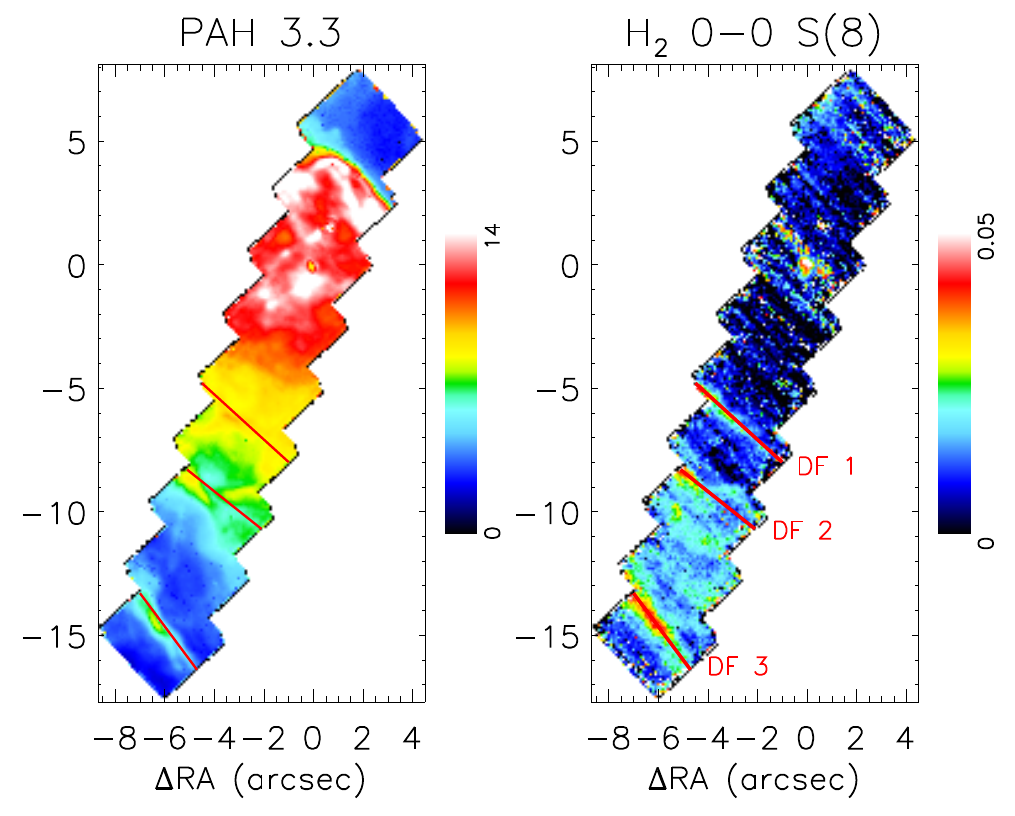}
    \caption{Intensity maps of the PAH 3.395\,$\mu$m, PAH 3.405\,$\mu$m, PAH 3.3\,$\mu$m and H$_2$\,0$-$0\,S(8) emission features in the Orion Bar region (1$^{\prime\prime}$=0.002\,pc). The three dissociation fronts (DF) are indicated in each map. Spaxel fluxes are in units of 10$^{-15}$\,ergs\,s$^{-1}$\,cm$^{-2}$. Typical errors on fluxes are in the range 10--20\%, except for H$_2$\,0$-$0\,S(8), where median error is 26\%.}
    \label{fig:orion_maps}
\end{figure*}

The Orion Bar is the outer region of the Orion Molecular Cloud, and is illuminated by the stars of Trapezium Cluster \cite{Peeters24}.  The Orion Bar observations with JWST$/$NIRSpec cover nine slightly overlapping pointings.
Detailed analysis of the NIRSpec data can be found in \cite{Berne22, Peeters24, Schroetter24, Chown24}. Adopting the standard distance of 414\,pc to the Orion Bar makes each 0\farcs1 spaxel equal 2$\times 10^{-4}$\,pc. 
We followed the same procedure as for NGC~6240 to fit the molecular hydrogen and ionized hydrogen emission lines, as well as the PAH features. The only exception was that there was no need to fit two kinematic components to each hydrogen emission line, so we only fit single Gaussians. The resulting maps are shown in Figure \ref{fig:orion_maps}.

\section{Theoretical PAH spectra}
\label{sec:DFT} 
To gain further insight into the 3.3 and 3.4\,$\mu$m PAH emission bands and, in particular, the nature of the double-peaked profile seen in the PAH 3.4\,$\mu$m feature,
here we examine DFT theoretical spectra of  molecules containing aliphatic sidebands. Following the methodology described in \citealp{rigo21} and \citealp{Kerkeni22} we compute the IR spectra of PAH molecules 
containing methyl, ethyl and vinyl side-groups. 
To keep the computations simple we chose Coronene, a molecule with a small number of carbons (N$_{\rm c}$=24) and computed DFT spectra of C$_{24}$H$_{12}$, C$_{24}$H$_{11}$-CH$_{3}$, C$_{24}$H$_{11}$-CH$_{2}$-CH$_{3}$ and
C$_{24}$H$_{11}$-CH=CH$_{2}$.
The computations presented here are not intended to be exhaustive but to give an indication of the strength and location of the PAH 3.4 $\mu$m feature in the presence of side-groups. We refer the reader to the works of \cite{yang20} and \cite{cabezas25} for a discussion of superhydrogenated and ring-containing aliphatic PAHs, respectively. 

\begin{figure}
	\includegraphics[width=1.0\columnwidth,angle=0]{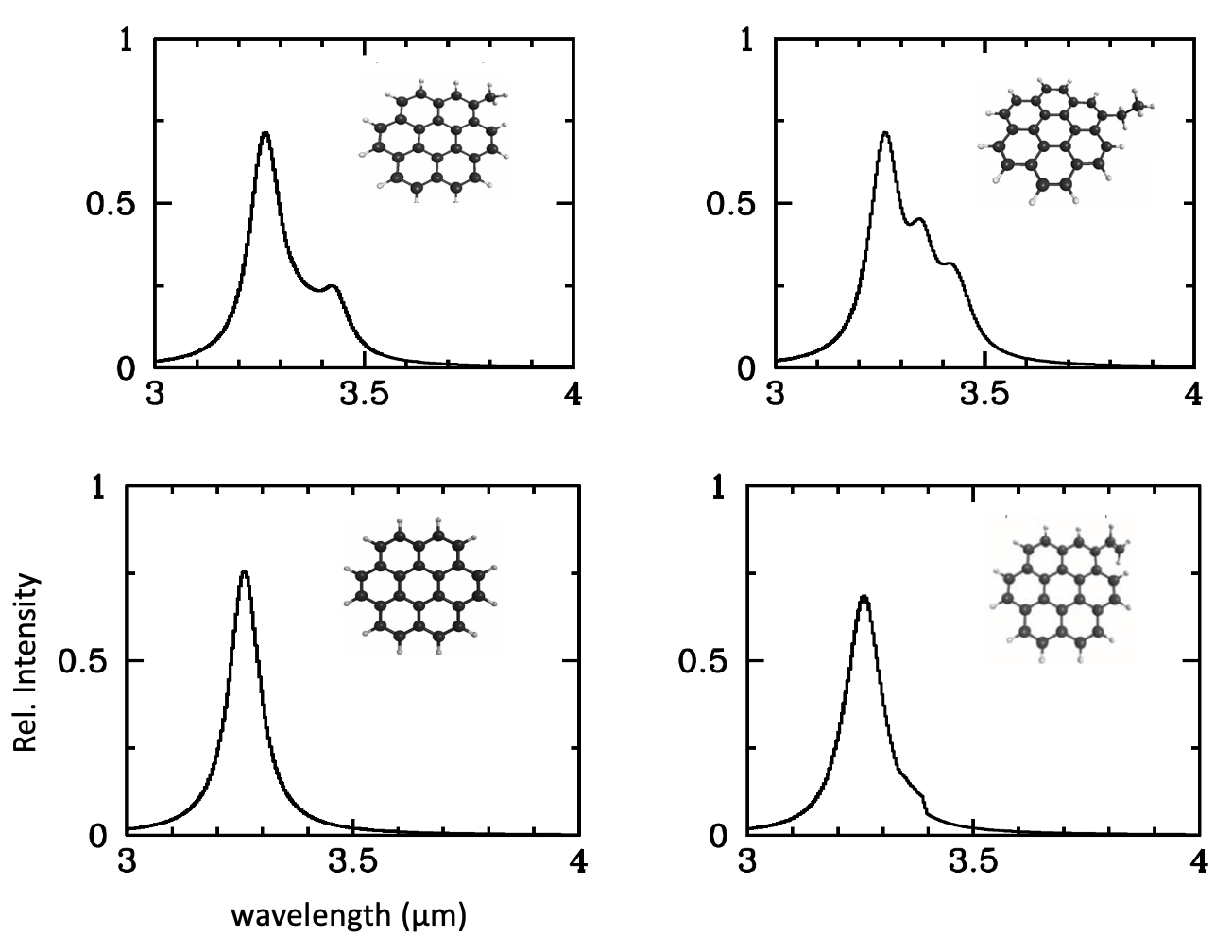}
    \caption{ DFT theoretical emission spectra of the 3-4 $\mu$m region showing aliphatic and aromatic C-H stretching vibrational modes for molecules with aliphatic side groups. From bottom left and going clockwise, Coronene C$_{24}$H$_{12}$, Coronene with methyl side-group C$_{24}$H$_{11}$-CH$_{3}$, Coronene with ethyl side-group C$_{24}$H$_{11}$-CH$_{2}$-CH$_{3}$, Coronene with vinyl side-group C$_{24}$H$_{11}$CH=CH$_{2}$.}
    \label{fig:dft}
\end{figure}

The structures and vibrational frequencies of the four molecules presented here were optimised with
B3LYP$/$6-311++G(d,p) using Gaussian 16\footnote{Gaussian 16, Revision C.01, M. J. Frisch et al., Gaussian Inc., Wallingford CT (2016).}.
Infrared spectra were computed from harmonic frequencies and intensities according to \cite{Bargaza25}.
In Figure \ref{fig:dft} we show the model emission spectra in the 3-4\,$\mu$m range for neutral PAHs exposed to photons of 6\,eV energy. We consider the full cascade model (for details see \citealp{rigo21} and references therein). It is evident from the spectra that the addition of ethyl$/$methyl side-groups shifts the peak of the 3.4\,$\mu$m feature redward towards the observed 3.405 $\mu$m peak. Likewise, a peak blueward of 3.4 $\mu$m becomes prominent as the number of CH$_{3}$ side-groups decreases. 

The bond dissociation energies (BDE) to remove the attached
aliphatic side groups are 4.14, 3.94, and 4.51 eV for C$_{24}$H$_{11}$-CH$_{3}$, C$_{24}$H$_{11}$-CH$_{2}$-CH$_{3}$, and C$_{24}$H$_{11}$-CH=CH$_{2}$, respectively (e.g., \cite{Buragohain20}). 
In the harsh
conditions of an astronomical source, the chemical pathways
might be different and perhaps there might be an intermediate product before the molecule 
completely loses the attached side groups. 
Of the three molecules considered here, vinyl side groups ($-$CH$=$CH2) are more strongly bound
and 4.51 eV is needed to remove the 
group from the parent molecule. 

\begin{figure*}
	\includegraphics[width=1.0\columnwidth,angle=0]{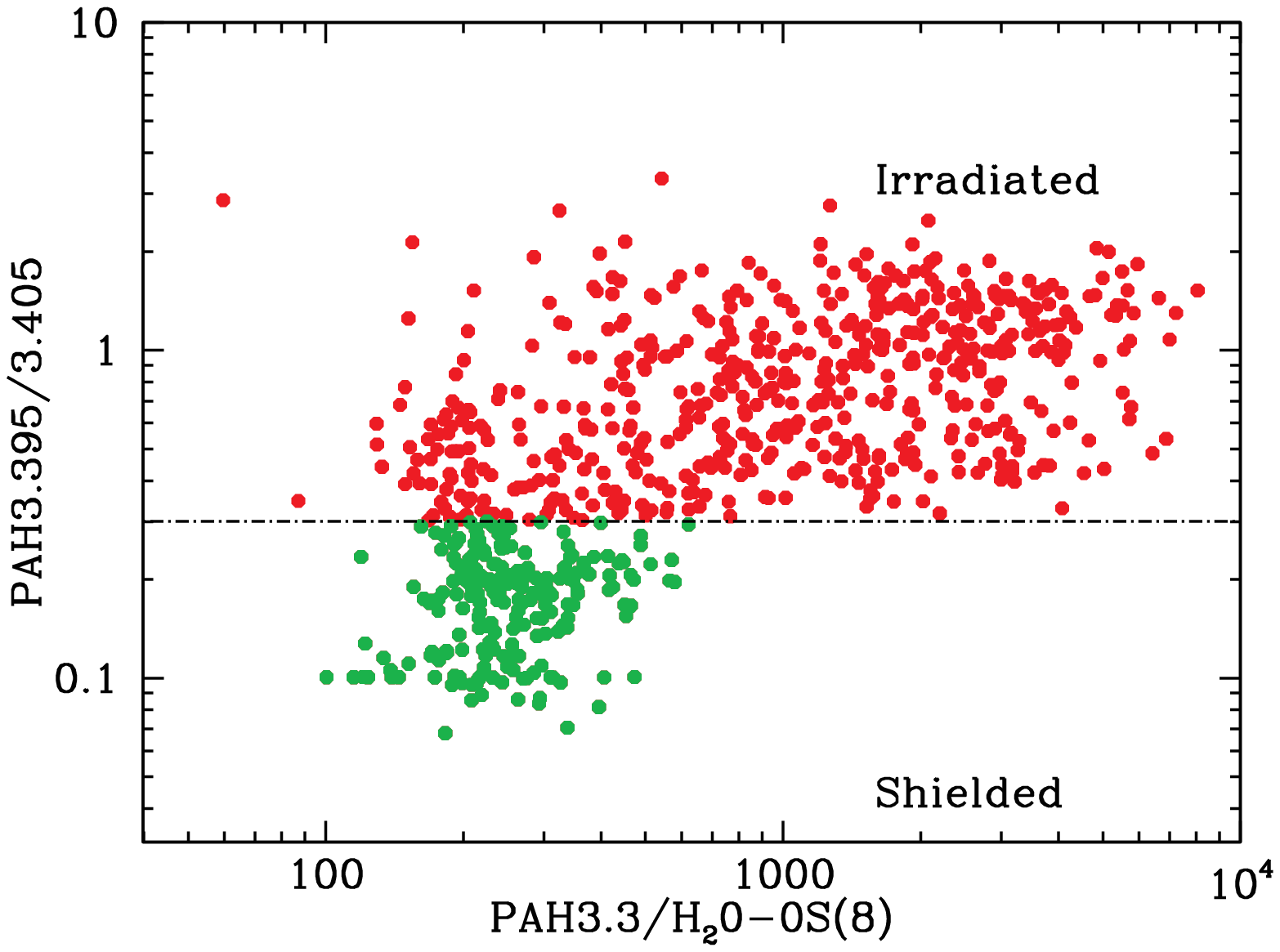}
   \includegraphics[width=1.0\columnwidth,angle=0]{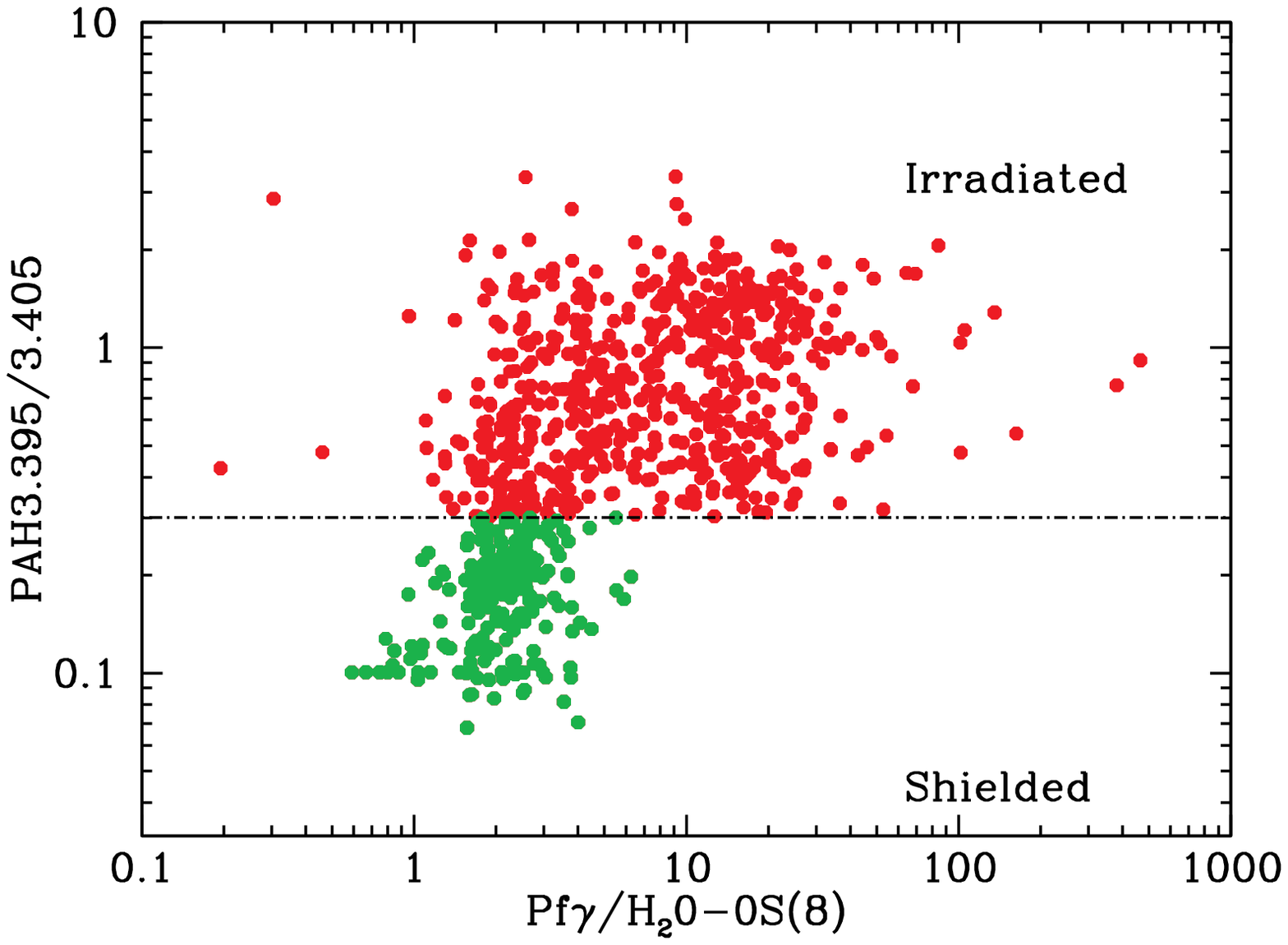}
    \caption{PAH 3.395$/$3.405 as function of PAH3.3$/$H$_{2}$ flux ratio for the rotational 
    H$_{2}$ 0-0 S(8) emission line at $\lambda$ = 5.05 $\mu$m (left panel) and the Pfund${\gamma}/$H$_{2}$ ratio  (right panel) for the Orion Bar. We assign those points with PAH3.395$/$3.405<0.3 to ``shielded'' PAHs (green circles) and the remaining points with values above this ratio as ``iraddiated'' PAHs (red circles), as discussed in the text.}
    \label{fig:pointcloud3p3overh2}
\end{figure*}

\section{\texorpdfstring{Do H$_{2}$ molecular clouds Shield PAH molecules?}{Do H2 molecular clouds Shield PAH molecules?}}

In this Section we set out to investigate how the physical ISM conditions impact the presence and strength of the two components that make up the PAH 3.4\,$\mu$m feature. We want to establish what sets the overall strength of the two components seen in Figure \ref{fig:spaxelspec}. In the pre-JWST era the double-peaked nature of the feature was not known and investigations were focused on the relation between aliphatic and aromatic stretches, which correspond to the PAH 3.4\,$\mu$m and PAH 3.3\,$\mu$m emission features, respectively. For instance, \cite{Yang23} investigated the ratio PAH~3.3/PAH~3.4 in samples of Planetary Nebulae (PNs), Reflection Nebulae (RNs) and Photo-Dissociation Regions (PDR) 
and concluded that the aliphatic feature is stronger in UV-poor environments (they treat the PAH~3.4\,$\mu$m as a single feature). Nonetheless, the Red Rectangle, a PN illuminated by HD 44179 of T$_\mathrm{eff} \sim$7750 K, had a rather low PAH 3.4\,$\mu$m intensity, much lower than that of PDRs and RNs illuminated by stars of much higher T$_\mathrm{eff}$ (\citealp{Geballe85}).  This implies that although the hardness of the radiation field is one of the parameters affecting the survival of aliphatic side-groups attached to PAHs it is probably not the only one.

\subsection{The case of the Orion Bar}
We employ JWST spatially resolved data to further investigate which factors determine the 
relative strength of the two PAH 3.4 sub-features at 3.395 and 3.405\,$\mu$m.
For this purpose, we first turn to the JWST$/$NIRSpec IFU spatially resolved observations of the Orion Bar since its exquisite resolution enables a detailed investigation of the spatial extent of the features.
Figure \ref{fig:orion_maps} shows the spatially resolved maps of the Orion Bar in the  
PAH3.3, PAH3.395, PAH3.405 and H$_{2}$ 0-0 S(8) lines. Similar maps have been presented in \cite{Peeters24, Schroetter24} but for consistency we show our own analysis here. 
\begin{figure}
\includegraphics[width=1.0\columnwidth]{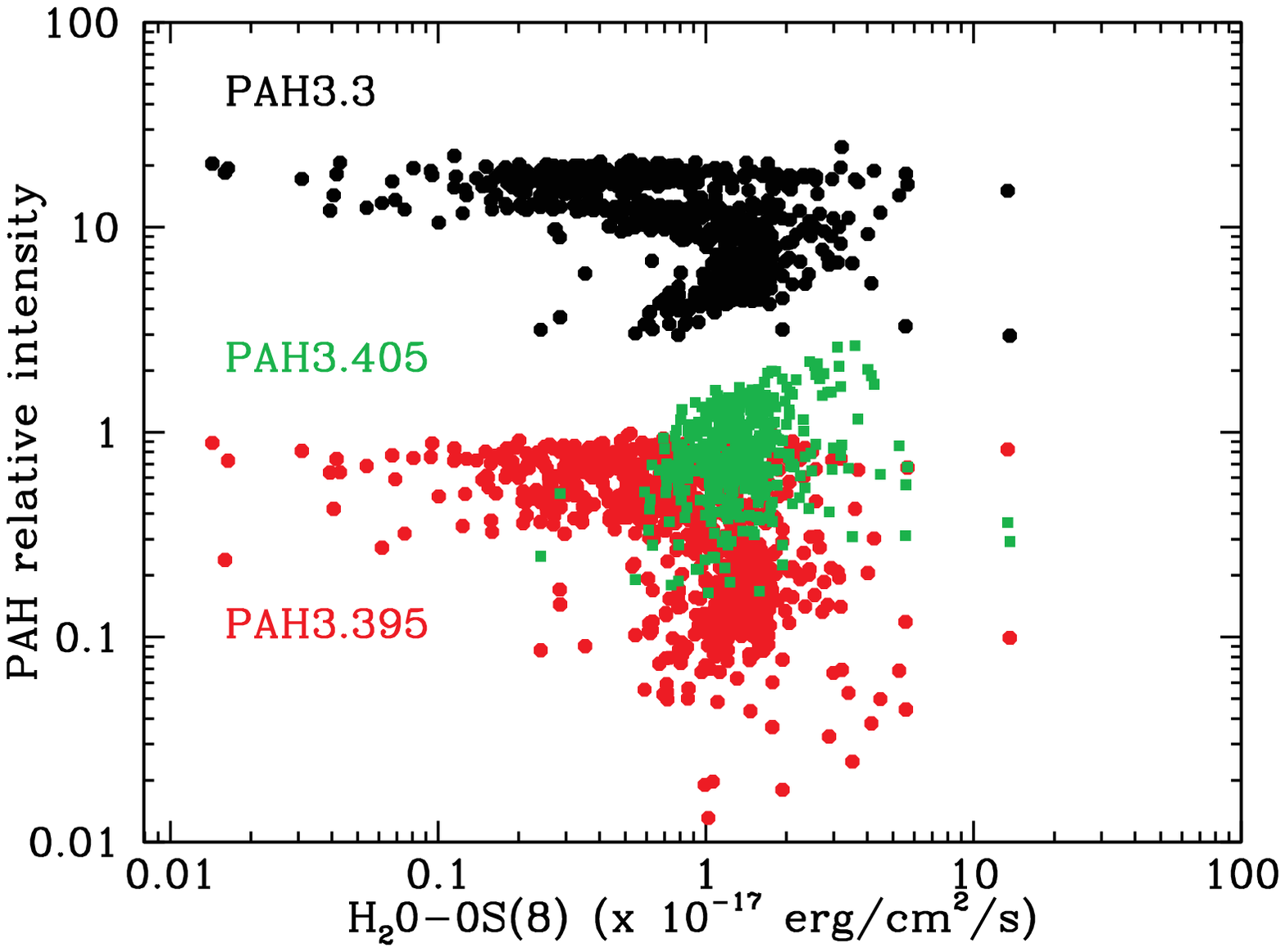}
    \caption{The distribution of the spatially resolved points for the PAH 3.3 (black), 3.395 (red) and 3.405 (green) $\mu$m features as a function of the H$_{2}$ 0-0 S(8) line, for the Orion Bar.   }
    \label{fig:pointcloudh2}
\end{figure}
Inspection of the maps in Figure \ref{fig:orion_maps} shows that the PAH 3.405\,$\mu$m feature
peaks in regions where the H$_{2}$ 0-0 S(8) emission is strong, hinting at a possible link between the two. A prominent enhancement in the strength of PAH 3.405 feature is seen at the southern end of the mapped region, coincident with the Dissociation Fronts (DF 2 and 3) discussed in \cite{Habart23, Peeters24} and indicated in our maps as well.
Emission from PAH 3.395, on the other hand, dominates in regions which are closer to the Trapezium stars, located towards the north-west (off the map, roughly in the direction of the long axis of the image mosaic). The morphology of the PAH 3.395\,$\mu$m emission broadly follows that of the PAH 3.3\,$\mu$m,
the latter arising from aromatic C$-$H stretches which are more robust against destruction by UV photons. 
The spatial distribution of the two features, therefore, implies that the PAH 3.405 likely arises in regions which are shielded from intense UV radiation, with the molecular H$_{2}$ providing the shielding. 

A link between PAH emission morphology and that of other dense gas tracers has already been suggested by various studies. \cite{rigo02, roussel07, mps10} demonstrate that a strong correlation exists between PAH and the pure H$_{2}$ rotational lines. More recently, \cite{AAH20} investigated the relation between the detection of the PAH 11.3\,$\mu$m feature in the nuclear regions of Seyfert galaxies and the properties of the cold molecular gas. They found that the strength of PAH 11.3 emission correlates positively with the column density implying that the molecular gas likely plays a role in shielding the PAH molecules in the harsh environments of Seyfert nuclei.
The PAH 3.395, on the other hand, traces regions which have been exposed to UV photons as has been suggested by e.g. \cite{Schroetter24}.

\begin{figure*}
	\includegraphics[width=0.9\textwidth,angle=180]{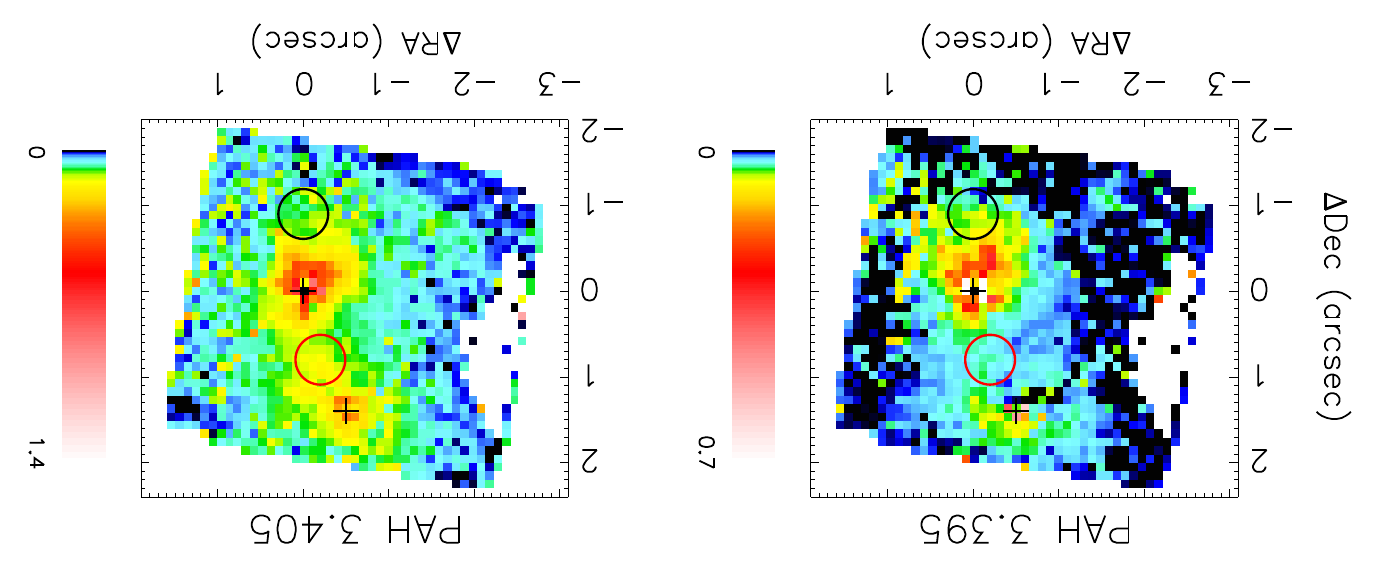}
   \includegraphics[width=0.9\textwidth,angle=180]{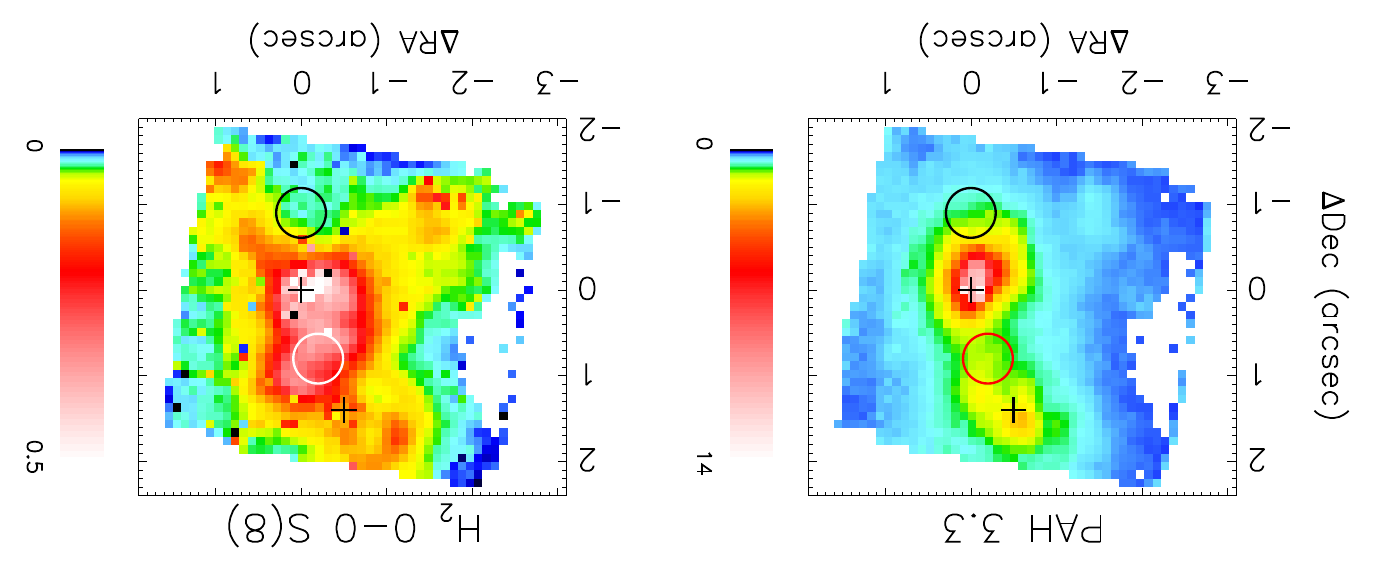}
    \caption{Intensity maps of the PAH 3.395\,$\mu$m, PAH 3.405\,$\mu$m, PAH 3.3\,$\mu$m and H$_2$\,0$-$0\,S(8) emission features in NGC6240 (1$^{\prime\prime}$=524\,pc) . The black crosses denote the location of the two nuclei. The red (white for the H$_2$ 0-0 S(8) map) and black circles represent the interaction zone and the region south of the southern nucleus as discussed in the text. Spaxel fluxes are in units of 10$^{-15}$\,ergs\,s$^{-1}$\,cm$^{-2}$. Typical errors on fluxes are in the range 10--20\%.}
    \label{fig:n6240maps}
\end{figure*}

In Figure \ref{fig:pointcloud3p3overh2} (left) we plot the PAH3.395/PAH3.405 ratio as a function of the PAH3.3$/$H$_{2}$0$-$0 S(8) ratio for the Orion Bar (the H$_{2}$ pure rotational line at 5.053\,$\mu$m). 
The majority of the spatially resolved points appear to have PAH3.395$/$PAH3.405 values above $\sim$0.3 with a median of around 1.
Interestingly, there appears to be a cluster of points with low values in the PAH3.395$/$3.405 ratio, corresponding to PAH molecules with a notably stronger PAH3.405 feature. 
These points all have PAH3.3$/$H$_{2}$ in the range $\sim$100--750. The upper value of this ratio together with a value for the PAH3.395$/$3.405 ratio of $\sim$0.3
mark a sharp transition into a region which is devoid of any observational points. We interpret the presence of this clear locus of points with PAH3.395$/$3.405$<$0.3 
as evidence of shielding of the more fragile aliphatic PAH molecules, responsible for the 3.405\,$\mu$m peak, by clouds of molecular hydrogen. As the PAH3.3$/$H$_{2}$ ratio grows bigger (hence the H$_{2}$ emission becomes weaker), exposure to FUV photons will tend to destroy some of the fragile aliphatic side chains responsible for the PAH 3.405 emission.

In order to locate the region(s) where PAH~3.395$/$PAH~3.405$<$0.3 we take the ratio of the PAH~3.395 and PAH~3.405 maps and apply a cut of 0.3 to the ratio values. In this manner, we create a mask which is shown in Figure \ref{orion_mask}. The areas with PAH~3.395$/$PAH~3.405$<$0.3 (shown in green) are coincident with peaks of the H$_{2}$ 0-0 S(8) emission (Figure \ref{fig:orion_maps}) which shows higher intensity in the lower half of the Orion mosaic. These peaks in H$_{2}$ emission trace the DF fronts (e.g. \citealp{Habart23, Peeters24}). In these regions, dissociating FUV photons are attenuated enough to just excite the aliphatic molecules without destroying them.
As one moves north of the molecular region past the DF regions, and toward the Trapezium stars, higher UV photon fluxes facilitate the processing of the aliphatic molecules, removing some of the alkylated bands and resulting in a gradual increase in the PAH~3.395$/$PAH~3.405 ratio. This is also evident in Figure \ref{fig:pointcloudh2} where we plot the spatially resolved points for each of the three PAH features as a function of the strength of the molecular hydrogen line  H$_{2}$(0$-$0)S8. It is clear that the PAH~3.405 points cluster around high values of H$_{2}$ and whilst the molecules responsible for the PAH~3.395 peak are also present, their emission is weaker. In addition, it appears that the PAH 3.405 intensity increases with increasing H$_2$ intensity, contrary to the behaviour of both the PAH 3.395 and PAH 3.3 features.

A similar picture emerges when we examine the variation of the PAH sub-features as a function of the ionizing flux (in this case parameterized by the Pf$_{\gamma}$ line). The right-hand plot of Figure \ref{fig:pointcloud3p3overh2} shows PAH~3.395$/$PAH~3.405 as a function of the Pf$_{\gamma}$/H$_{2}$ ratio.
We observe a similar distribution of the spatially resolved points of the Orion Bar as we saw earlier (in the left-hand side plot of the same figure).
We note the existence of a clear cluster of points with values of the ratio of 
PAH~3.395$/$PAH~3.405$<$0.3 and Pf$_{\gamma}/$H$_{2}<$10. As before, in these regions the molecular H$_{2}$ clouds are shielding the more fragile aliphatic molecules from ionizing photons (as is evident from the low values in the Pf$_{\gamma}/$H$_{2}$ ratios). Once the Pf$_{\gamma}/$H$_{2}$ ratio crosses the threshold value of 10, we move to more irradiated regions where the UV field is stronger and where the H$_{2}$ molecules are photo-dissociated. In these regions, photo-processing of aliphatic PAHs is likely to take place resulting in large PAH~3.395$/$PAH~3.405 ratios. The analysis of the Orion Bar data has enabled us, for the first time, to  locate regions where fragile PAH molecules are shielded from FUV photons. We were able to trace these regions through the weak 3.405 $\mu$m emission of the PAH molecules.  

PAH molecules in regions exposed to a hard radiation field exhibit weak 3.405\,$\mu$m emission, presumably because the fragile aliphatic side chains are destroyed by the energetic FUV photons. These regions are labelled as ``irradiated'' regions by \cite{Schroetter24}. The presence of dense molecular gas, traced by the H$_2$ emission, shields the PAH molecules from the energetic FUV photons. This not only softens the radiation field, but also weakens it, so that there is less PAH emission overall. However, as the PAH molecules remain intact, they exhibit all three emission features, at 3.3, 3.395 and 3.405\,$\mu$m, with PAH~3.405 stronger than PAH~3.395 emission.

\subsection{The case of NGC~6240}
We now turn to NGC~6240 as we seek to establish whether the same mechanism, where the molecular H$_{2}$ is shielding the PAHs from strong UV fields is observed on larger scales in nearby galaxies with strong molecular hydrogen emission, even though the spatial resolution for the NGC\,6240 observations is several orders of magnitudes coarser than for the Orion Bar.
Figure \ref{fig:n6240maps} shows the intensity maps of the PAH~3.3,
PAH~3.395, PAH~3.405 and H$_{2}$ 0-0 S(8) lines in the nuclear region of NGC\,6240. The location of the two NGC\,6240 nuclei is indicated with crosses on all the images. The morphology of the PAH~3.3 map delineates the regions in the galaxy where the PAH molecules are excited by FUV photons. Within these regions, we see a strong enhancement of the PAH~3.405 emission relative to the PAH~3.395 emission in the so-called \emph{interaction} region, marked by a red circle in Figure \ref{fig:n6240maps}.
Conversely, little PAH~3.405 emission is seen from the region south of the southern nucleus, indicated by the black circle. The corresponding region in the H$_2$ emission map shows a "hole", whereas considerable PAH~3.395 emission is present. The location of this "hole" coincides with the ionized bubble south of the southern nucleus, as identified and discussed in \citealp{Ceci25}.

The data points with values of PAH~3.395$/$PAH~3.405$<$0.3 (the "shielded" region) are all located in the "interaction region" between the two nuclei of NGC~6240. This is coincident with the region where the H$_{2}$ molecular hydrogen emission is strong, as has already been noted by \cite{Tecza00, Lutz03, Ceci25, Hermosa25} and is evident in the map shown in Figure \ref{fig:n6240maps}. This is also consistent with the findings from the analysis of the Orion Bar where the peak of the PAH~3.405 intensity is coincident with regions of high H$_{2}$ intensity as shown in Figure \ref{fig:pointcloudh2}. 
Our qualitative interpretation uses the H$_{2}$\,0$-$0\,S(8) as an indicator of the presence of warm molecular gas, given its proximity in wavelength to the PAH 3.3 features.  Although the higher 0$-$0 S transitions trace warmer gas, we note that in NGC 6240 all the H$_2$\,0$-$0 transitions exhibit the same morphology, with a strong peak at the interaction region, in between the two nuclei (Hermosa Mu{\~n}oz et al. in prep.).  We will revisit this issue in the Discussion section.

\begin{figure*}
	\includegraphics[width=1.0\columnwidth,angle=0]{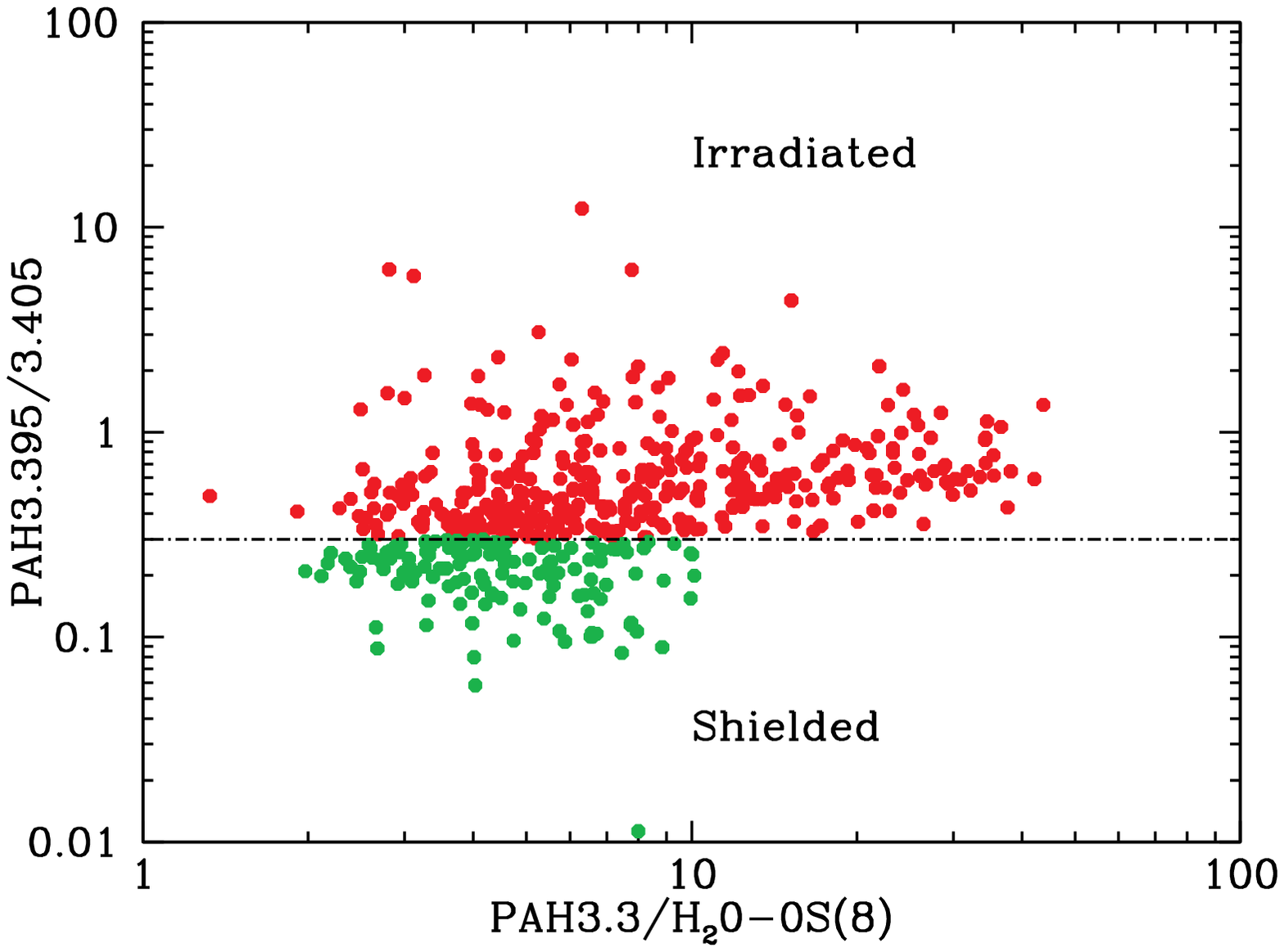}
   \includegraphics[width=1.0\columnwidth,angle=0]{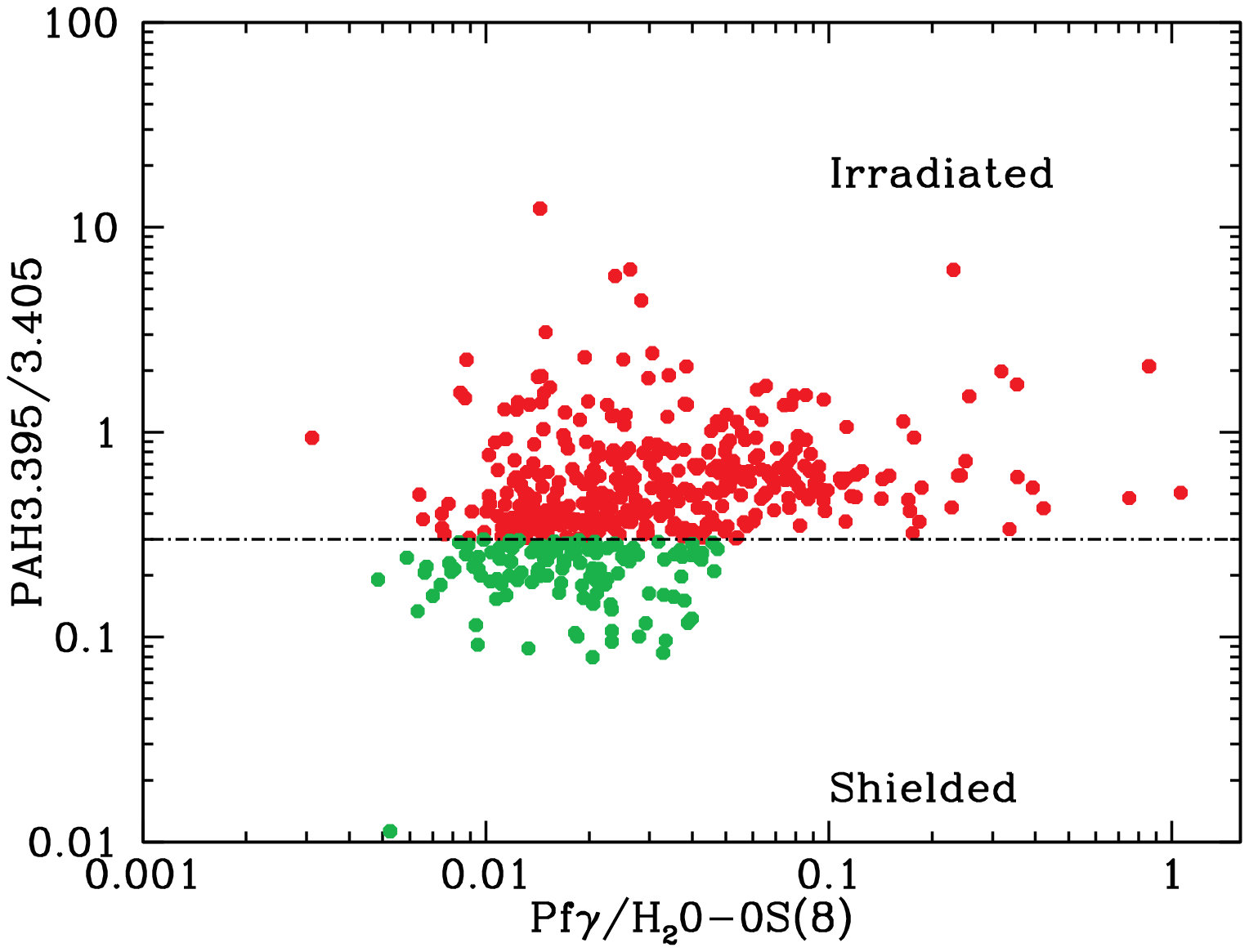}
    \caption{PAH 3.395/3.405 as function of PAH3.3/H2 flux ratio for the rotational H$_{2}$0-0 S(8) emission line at $\lambda$= 5.05 $\mu$m (left panel) and the Pfund${\gamma}/$H$_{2}$
ratio (right panel), for NGC 6240. We assign those points with PAH3.395/3.405<0.3 to “shielded” PAHs (green circles) and the remaining points with values above this ratio
as “irradiated” PAHs (red circles), as discussed in the text. }
    \label{fig:n6240points}
\end{figure*}

NGC\,6240 is well-known for being H$_{2}$-luminous not only in ro-vibrational lines (e.g. \citealp{Joseph84, Tecza00, vdWerf93}) but also in pure rotational lines. \cite{Lutz03} found that the luminosity of the H$_{2}$ pure rotational lines of 1.8$\times$10$^{9}\,$L$_{\odot}$ corresponds to 0.3\% of the bolometric luminosity of the galaxy. 
In Figure \ref{fig:n6240points} we plot the ratio of the PAH3.395/PAH3.405 as a function of the PAH3.3$/$H$_{2}$0$-$0 S(8) ratio. The NGC\,6240 points broadly follow the same 
distribution as with the Orion Bar data (Figure \ref{fig:pointcloud3p3overh2} left).
A fraction of the spatially resolved NGC\,6240 data points have low PAH~3.395$/$PAH~3.405 values ($<$0.3), these points are indicated with the green symbols in Figure \ref{fig:n6240points}.
Interestingly, all NGC~6240 data points with low PAH~3.395$/$PAH~3.405 values have PAH3.3$/$H$_2$ values below 10, and as was the case with the Orion Bar, the bottom right-side portion of the plot is devoid of points. Once again, we see evidence of H$_{2}$ molecular gas shielding: the large amount of molecular hydrogen present in NGC~6240 shields the aliphatic PAHs from energetic FUV photons. Following the same procedure as outlined earlier, we create a `mask' for NGC\,6240 by dividing the two maps PAH~3.395$/$PAH~3.405 and applying the same cut of 0.3 for the ratio. The mask is shown in Figure \ref{n6240_mask}, with the two circles marking the ``shielded'' region and the ``irradiated'' region overlaid. It is very clear that the PAH~3.405 emission is strong in the shielded region, and weak in the irradiated region, in comparison to the PAH~3.395 emission.

 Figure \ref{fig:n6240points} (right)
shows the ratio of PAH~3.395$/$PAH~3.405 as a function of the Pf$_{\gamma}/$H$_2$0$-$0 S(8) ratio.
The shape of the distribution of the NGC~6240 points loosely follows that of the Orion Bar but with noticeably more scatter. This is not surprising given that in 
NGC~6240 the Pf$_{\gamma}$ emission is overall stronger and, in particular, around the two nuclei (see e.g., \citealp{Ceci25}). 
Whereas for the Orion Bar there are no points with PAH~3.395$/$PAH~3.405$<$0.3 and Pf$_{\gamma} /$H$_2 >$10, in NGC~6240 the H$_2$ emission lines are almost two orders of magnitude stronger. For the same PAH~3.395$/$PAH~3.405 ratio, the region devoid of points has Pf$_{\gamma} /$H$_2 >0.05$. As in the case of the Orion Bar, the PAH~3.395$/$PAH~3.405 ratio increases in regions where the ionizing flux (as traced by Pf$_{\gamma}$) is strong. 

It is also worth pointing out that the nominal spatial resolution of the NGC\,6240 observations is a lot coarser than that of the Orion Bar, with each spaxel corresponding to 52~pc in linear scale. Therefore, each spaxel in NGC\,6240 is sampling much larger areas including a mix of H$_{2}$ and PDR regions. The much stronger H$_{2}$ emission in NGC\,6240, enhanced by the presence of powerful shocks (e.g. \citealp{Sugai97, Lutz03}), causes the PAH3.3$/$H$_{2}$ and Pf$_{\gamma}/$H$_{2}$ ratios to have lower values than in Orion such that a direct comparison of the ratio ranges is not possible. We also note that the H$_2$ emission morphology is quite distinct to the PAH emission, as the excitation mechanisms are not the same. The analysis presented here is limited to the regions where PAH~3.3 emission is observed.

Still we note that there are two regions in NGC~6240 that exhibit clear signatures consistent with our hypothesis: (1) the "interaction region" exhibits strong H$_2$ emission, and a low PAH~3.395$/$PAH~3.405 ratio, consistent with shielding by dense molecular gas, and (2) the region south of the southern nucleus, showing evidence of stronger UV flux (see figure 21 of \cite{Ceci25}), little molecular gas emission, and exhibits PAH~3.395 emission, but little or no PAH~3.405. These regions are indicated by red and black circles in Figure \ref{fig:n6240maps} respectively.

\section{Discussion}

\begin{figure}
	\includegraphics[width=0.9\columnwidth,angle=0]{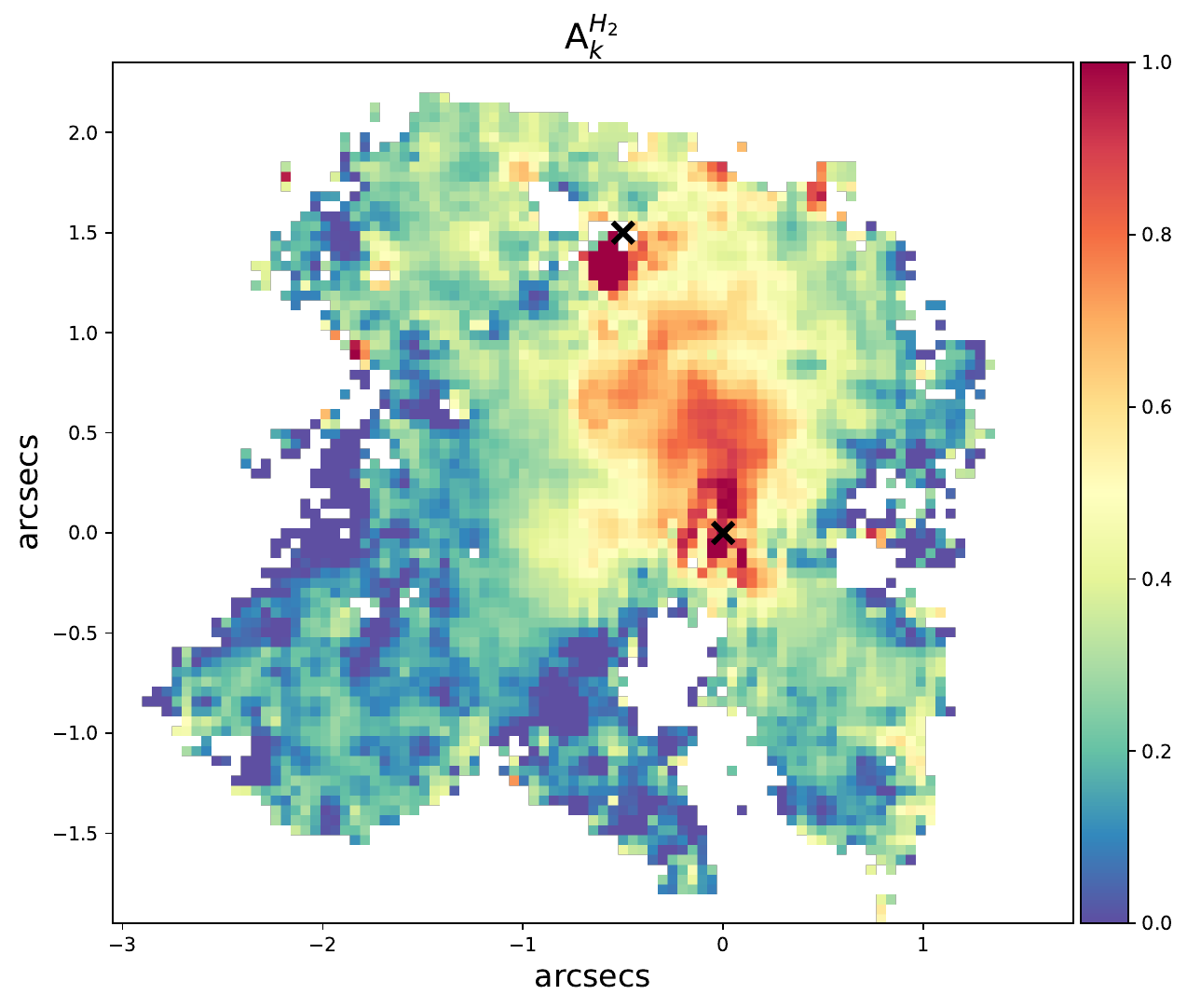}
    \caption{NGC 6240 K band extinction map (A$_{\mathrm{K}}$) based on H$_{2}$ rovibrational lines that share the same upper level (i.e., $\nu$, J.)}
    \label{fig:extinct}
\end{figure}
The 3.4 $\mu$m aliphatic C–H emission has been widely seen in various galactic astrophysical sources (e.g. \citealp{Sloan97, Mori14, Pilleri15,Boersma23}). However, the
detection of two separate features at 3.395 and 3.405 $\mu$m in external galaxies had not been observed prior to JWST. The two features have, so far, been seen in the Orion Bar data (\citealp{Peeters24}), and a number of other galactic sources (e.g.\citealp{Boersma23}). 
In their work, \cite{Schroetter24} employed template spectra to fit the 3.3 and the 3.4\,$\mu$m PAH bands by fitting single profiles to each of these two features. In contrast, in this work, we fit the two sub-features at 3.4\,$\mu$m separately with Drude profiles, and the 3.3\,$\mu$m feature using a combination of Drude and Gaussian profiles, as discussed in Section 2.1.2. Our method results in robust fits for every spaxel in the Orion Bar and NGC\,6240. For each of these two targets, through the analysis presented here, we have been able to identify regions where there appears to be an enhanced population of fragile aliphatic PAHs, which are responsible for the PAH~3.405 $\mu$m peak. Their survival appears to depend on the presence of dense molecular gas that shields the PAH molecules from energetic UV photons, whilst still allowing lower energy UV photons to excite the PAH molecules.

The comparison of the distribution of aromatic and aliphatic molecules (Figure \ref{fig:pointcloudh2}) reveals that the PAH~3.395 emitting molecules follow broadly the distribution of the aromatic ones whereas those responsible for the PAH~3.405 band appear to cluster around high intensity values of the H$_{2}$ 0$-$0 S(8) line. The theoretical DFT spectra show that the 3.405\,$\mu$m peak is prominent when there are -CH$_{3}$ side-groups attached to the PAH molecule. Given that the aliphatic-to-aromatic (PAH3.4$/$PAH3.3) ratio decreases with increasing intensity of the FUV radiation field (\citealp{Sloan97, Mori14}), it is possible that the molecules responsible for the PAH~3.405 $\mu$m peak will only survive when they are sufficiently shielded from FUV radiation. This is consistent with observations of an enhancement of PAH~3.405 when the H$_{2}$ intensity is high, as shown in Figure \ref{fig:pointcloudh2}.

Previous observations of PDRs have also shown that
emission from aromatic PAH molecules and that from H$_2$ excited by fluorescence 
are nearly co-spatial, with H$_2$ sometimes seen to extend slightly deeper into molecular
clouds (e.g. \citealp{Sellgren90, Tielens93, Habart03}). 
This co-spatial emission is not unexpected
since both species are excited by FUV photons.
PAH molecules can also be excited by lower energy
photons in the ultraviolet and optical, but with smaller absorption
cross sections (see \citealp{LiDraine01}); although FUV photons will always dominate the excitation whenever massive stars or AGN are present. However, H$_2$ is dissociated by FUV photons between 11.3 and 13.6\,eV
where it is not self-shielded, whereas aromatic PAHs survive the absorption
of these photons.

In the case of relatively dense molecular clouds, PAHs may reside on the surfaces of large grains. When the 
propagating photoionization$/$photodissociation front arrives at a
parcel of gas that was previously "dark cloud" material, PAHs will be ejected from the grains. This ejection
has two main consequences: the PAHs begin to emit following single-photon heating and second, the photoelectric heating rate increases, heating the H$_2$ gas (e.g., \citealp{Allers05}). In this scenario, H$_{2}$
emission should peak at slightly higher optical depth than aromatic bands but likely at the same optical depth as the aliphatic PAH~3.405 band.

Can we link the presence of the most fragile aliphatic PAHs, those responsible for the PAH~3.405 $\mu$m peak, to 
the physical conditions of the molecular clouds and in particular the levels of extinction required to shield the PAH molecules?
In the case of the Orion Bar, \cite{Peeters24} determined that the extinction towards DF3 has values of A$_\mathrm{V} \sim$ 1.8 to 2.30 mag assuming a screen or mixed dust scenario. Their analysis concluded that the variations seen in the profiles and the ratios of aliphatic to aromatic PAHs are also influenced by changes in the FUV radiation. 

A comparison of the maps shown in Figures \ref{fig:n6240maps} and \ref{n6240_mask} reveals that the ``shielded'' molecules in NGC\,6240 are located in the "interaction region" between the two nuclei. 
Figure \ref{fig:extinct} shows the A$_\mathrm{K}$ extinction map of NGC~6240, assuming a screen dust model. The map was created in a manner similar to that described by \cite{Rosenthal00}, where all detected H$_{2}$ vibrational transitions that share the same upper level ($\nu$, J) are used to determine the differential extinction (more details in Speranza et al., in prep). 
The highest values of the extinction are found  towards the southern nucleus of NGC~6240. This region has a value of A$_{\mathrm{K}}\sim$1.0 mag corresponding to an A$_{\mathrm{V}}\sim$10 mag. Extinction towards the interaction region is around A$_{\mathrm{V}}\sim$6 mag. The interaction region is the location where the aliphatic PAH ratio (PAH~3.395$/$PAH~3.405) is $<$0.3, hence this is where the "shielded PAHs" are located.
These PAHs must reside in the outer layers of the molecular clouds. The dense, inner regions of the molecular clouds likely correspond to very high levels of extinction that significantly attenuate the FUV photons, resulting in little or no PAH emission. 

Given that in the Orion Bar the extinction towards DF3 was estimated to be A$_{\mathrm{V}}\sim$1.8-2.30 mag (\citealp{Peeters24}), we conclude
that as long as the A$_\mathrm{V}$ values remains $\geq$2, aliphatic molecules are able to survive in the outer layers of molecular clouds, just past the transition zone between atomic and molecular hydrogen gas. It is, however, worth noting that although the extinction towards the Southern nucleus has the highest values, the radiation field is also very strong (e.g., \citealp{Ceci25}) potentially destroying the fragile PAH\,3.405 molecules. 

However, a major difference between the Orion Bar and NGC\,6240 is the source of H$_2$ excitation. While PDRs dominate the excitation of H$_2$ in the Orion Bar, shocks have been implicated as an extra mechanism for H$_2$ excitation in NGC\,6240, especially in the interaction region (e.g., \citealp{Tecza00, Lutz03}).  
An in-depth analysis of the impact of shocks on the PAH population in NGC\,6240 is beyond the scope of the present work and will be the focus of a forthcoming study.

\section{Conclusions}

The spatially resolved spectroscopic capabilities of JWST/NIRSpec IFU, combined with the telescope's unprecedented mid-IR sensitivity, has allowed us to clearly identify and analyze two sub-features in the PAH~3.4\,$\mu$m emission band, both in galactic and extra-galactic sources, for the first time. The PAH~3.4\,$\mu$m feature most often arises from bond stretches within the aliphatic side-chains attached to aromatic PAH molecules.  The two sub-features, the PAH~3.395\,$\mu$m peak and the PAH~3.405\,$\mu$m peak, have been observed in the Orion Bar region, with a very strong correlation between the PAH~3.405 emission and the molecular H$_2$ vibrational emission. The PAH~3.395 emission, on the other hand, follows the morphology of the much brighter PAH~3.3\,$\mu$m feature, arising from aromatic bond stretches.

Backed up by theoretical PAH spectra computed using DFT, we conclude that the PAH~3.405 sub-feature indicates the presence of fragile aliphatic side-chains of PAH molecules. The thermal emission is excited by UV photons of $\sim$5\,eV in energy. However, the presence of stronger UV radiation can cause these side-chains to completely break off. Thus, the presence of strong PAH~3.405 emission, in particular, values of the ratio PAH~3.395$/$PAH~3.405$<$0.3, corresponds to regions where the PAH molecules are {\it shielded} by dense molecular gas, so that only modestly energetic UV photons penetrate to excite the PAHs. 

Using spatially resolved plots of PAH~3.395$/$PAH~3.405 vs. PAH~3.3$/$H$_2$ on a  spaxel-by-spaxel basis, we show a clear demarcation between shielded and irradiated regions.  Strong PAH~3.405\,$\mu$m emission only arises in regions of high molecular gas density, evidenced by the complete lack of points with low PAH~3.395$/$PAH~3.405 and high PAH~3.3$/$H$_2$.  This is true both in the Orion Bar region, but also in NGC~6240, although the spaxels encompass very different size scales in the two sources.  In contrast, the PAH~3.395 sub-feature arises from more robust aliphatic side-chains that can withstand irradiation by more energetic UV photons, similar to the aromatic PAH molecules.  This {\it irradiated} aliphatic PAH~3.395 sub-feature displays an almost perfect correlation in its morphology with the PAH~3.3 emission (arising from aromatic PAH molecules), both in the Orion Bar, and in the nuclear region of NGC~6240. 

Dense molecular gas is required to shield the fragile aliphatic side-chains that give rise to the PAH~3.405 feature. However, if no UV photons can penetrate, the PAHs are not excited, and do not exhibit thermal emission. Thus, a "Goldilocks" level of shielding, corresponding to A$_\mathrm{V}$ of $\sim$a few, appears to provide the optimal conditions to observe both components of the PAH~3.4\,$\mu$m feature, leading to a double-peaked emission line. We conclude that the PAH~3.405\,$\mu$m and PAH~3.395\,$\mu$m emission features can provide robust diagnostics of the physical conditions of the ISM, and can be exploited to trace the energetics of the photon field penetrating molecular clouds.

\section*{Acknowledgements}
We thank the anonymous referee for perceptive and insightful comments which helped improve the manuscript.
NT and DR acknowledge support from STFC through
grant ST/W000903/1. FD acknowledges support from STFC through studentship ST/W507726/1. IGB is supported by the Programa Atracci\'on de Talento Investigador ``C\'esar Nombela'' via grant 2023-T1/TEC-29030 funded by the Community of Madrid.
AAH and LHM acknowledge financial support by grant PID2021-124665NB-I00 funded by the Spanish Ministry of Science and Innovation and the State Agency of Research MCIN/AEI/10.13039/501100011033, PID2021-124665NB-I00 and ERDF A way of making Europe.
MPS acknowledges support under grants RYC2021-033094-I, CNS2023-145506 and PID2023-146667NB-I00 funded by MCIN/AEI/10.13039/501100011033 and the European Union NextGenerationEU/PRTR.

Computational resources were provided by the Advanced Research Computing (ARC) facility at the University of Oxford [1]. We gratefully acknowledge the use of ARC for supporting the quantum chemical calculations performed in this work.

This work is based on observations made with the NASA/ESA/CSA James Webb Space Telescope. The data were obtained from the Mikulski Archive for Space Telescopes at the Space Telescope Science Institute, which is operated  by the Association of Universities for Research in Astronomy, Inc., under  NASA contract NAS 5-03127 for JWST; and from the European JWST archive (eJWST) operated by the ESAC Science Data Centre (ESDC) of the European Space Agency. These observations are associated with programs  \#1328 and \#1670.

\section*{Data Availability}
 
The JWST data used in this work are publicly available as part of DD-ERS Program 1288 (PI: O. Berne), and GTO1 - Program  1265 (PI: A. Alonso-Herrero) and downloadable from the MAST archive.




\bibliographystyle{mnras}
\bibliography{ThatteBib} 







\appendix

\section{Shielded Regions}

For the Orion Bar and NGC\,6240 observations, We present "masks" that show the regions where the PAH\,3.395/PAH\,3.405 ratio is <0.3, corresponding to regions where the PAH molecules are "shielded" from energetic UV photons by dense molecular gas.
\begin{figure*}
	\includegraphics[width=0.9\columnwidth,angle=0]{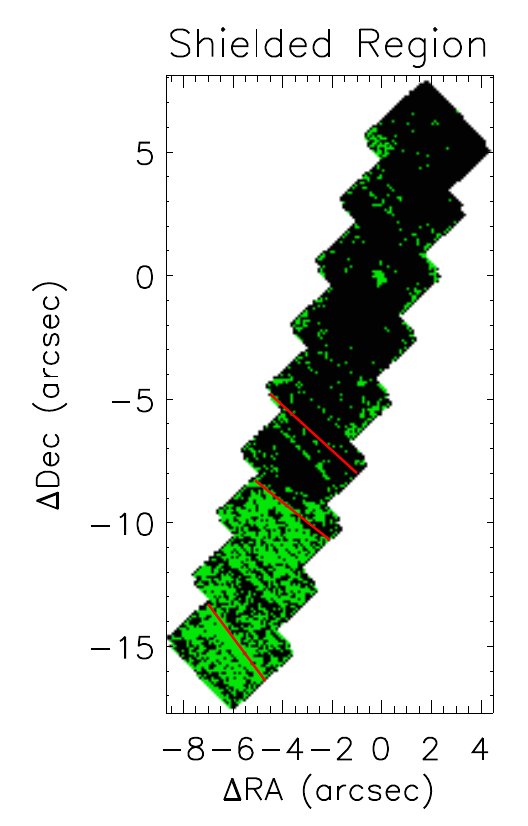}
\caption{The Orion Bar mask, green areas represent spaxels with PAH 3.395$/$3.405 values of $<$0.3 corresponding to "shielded" PAHs, as discussed in the text. The red lines mark the dissocation fronts.}
    \label{orion_mask}
\end{figure*}

\begin{figure*}
	\includegraphics[width=0.9\columnwidth,angle=90]{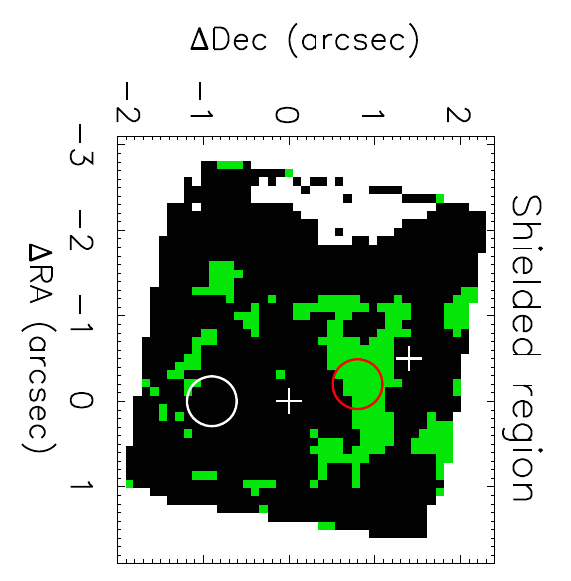}
\caption{The NGC 6240 mask, green areas represent spaxels with PAH 3.395$/$3.405 values of $<$0.3 corresponding to "shielded" PAHs, as discussed in the text. The white crosses mark the locations of the two nuclei. The two circles mark the same regions as in Figure \ref{fig:n6240maps}, except that the black circle has become white, to make it visible.  The red circle corresponds to a ``shielded'' region, whereas the white circle marks an ``irradiated'' region.}
    \label{n6240_mask}
\end{figure*}

\vfill

\bsp	
\label{lastpage}
\end{document}